\pdfoutput=1
\documentclass[acmtog,noacm]{acmart}
\renewcommand\footnotetextcopyrightpermission[1]{}
\setcopyright{none}                

\usepackage{amsmath}
\usepackage[utf8]{inputenc}
\usepackage{siunitx}
\usepackage{graphicx}
\usepackage[nolist]{acronym}

\usepackage[linesnumbered,ruled,vlined]{algorithm2e}

\definecolor{twred}{rgb}{0.85,0,0.2}
\definecolor{twgreen}{rgb}{0,0.65,0.2}

\def\clap#1{\hbox to 0pt{\hss #1\hss}}%
\def\initials#1{\protect\clap{\smash{\raisebox{1.4ex}{\tiny{\textsf{\textit{~~#1}}}}}}}%
\makeatletter
\newcommand{\NOTE}[3]{\protect\@ifundefined{hidecomments}{%
  \strut{\color{#2}{\hspace{0pt}\initials{#1}\protect{{\small$\lfloor$}#3{\small]}}}}%
  }{}}
\newcommand{\EDITbyauthor}[4][]{\protect\@ifundefined{hidecomments}{%
  \strut{\color{#3}{\hspace{0pt}\initials{#2}\protect\sout{#1}{#4}}}%
  }{}}
\newcommand{\EDITredandgreen}[4][]{\protect\@ifundefined{hidecomments}{%
  \strut{\color{twred}{\hspace{0pt}\protect\sout{#1}{\color{twgreen}{#4}}}}%
  }{}}
\newcommand{\EDITgreenonly}[4][]{\protect\@ifundefined{hidecomments}{%
  \strut{\color{twgreen}{#4}}%
  }{}}

\newcommand{\EDIT}[4][]{\EDITbyauthor[#1]{#2}{#3}{#4}}

\newcommand{\NOTEboxed}[3]{\protect\@ifundefined{hidecomments}{%
  {\centering\fbox{\parbox{0.97\linewidth}{\protect\EDIT{#1}{#2}{#3}}}}%
  }{}}

\def\myhrulefill{\leavevmode\leaders\hrule height 2pt\hfill\kern\z@}

\makeatother

\def\ignore#1{}

\usepackage{bm}

\newcommand{\IR}{\ensuremath{\mathbb{R}}}

\makeatletter
\newcommand{\currentfontsize}{\f@size}
\makeatother

\usepackage{cleveref}
\crefname{section}{Sec.}{Secs.} %
\Crefname{section}{Sec.}{Secs.} %
\crefname{figure}{Fig.}{Figs.} %
\Crefname{figure}{Fig.}{Figs.} %

\usepackage{xcolor}

\SetCommentSty{algcomment}

\begin{document}

\title{Feature-Guided Diffusion for Non-Dif\-fer\-en\-tia\-ble Inverse Rendering}
\author{Andrei-Timotei Ardelean}
\affiliation{%
  \institution{Friedrich-Alexander-Universität}
  \city{Erlangen-Nürnberg}
  \country{Germany}}
\email{timotei.ardelean@fau.de}

\author{Michael Fischer}
\affiliation{%
  \institution{Adobe Research}
  \city{London}
  \country{United Kingdom}}
\email{mifischer@adobe.com}

\author{Tim Weyrich}
\affiliation{%
  \institution{Friedrich-Alexander-Universität}
  \city{Erlangen-Nürnberg}
  \country{Germany}}
\email{tim.weyrich@fau.de}

\author{Tomáš Iser}
\affiliation{%
  \institution{Charles University, Faculty of Mathematics and Physics}
  \city{Prague}
  \country{Czech Republic}}
\email{tomas.iser@matfyz.cuni.cz}

\begin{acronym}
\acro{CMA}{Covariance matrix adaptation}
\acro{CMA-ES}{Covariance matrix adaptation evolution strategy}    
\end{acronym}

\begin{abstract}
Inverse rendering is traditionally solved via differentiable renderers and gradient descent, which requires substantial problem-specific engineering and is prone to getting stuck in local minima due to ambiguities.
Derivative-free approaches alleviate engineering requirements, but often heavily depend on a good problem initialization.
In this work, we propose Feature-Informed Diffusion Evolution (FIDE), a fully black-box framework that requires no gradients or specific initialization: the renderer is treated as an opaque function whose only requirement is to produce images. 
Our key insight is \emph{feature guiding}: rather than reducing each candidate rendering to a scalar loss value, we use a Vision Transformer (ViT) to extract dense visual features from it.
We subsequently use these features to train a diffusion-based candidate proposal model, allowing the network to use visual cues to predict parameters that would match the target image. 
The candidate solutions proposed by this diffusion model are then refined in a closed loop with a CMA evolution strategy, continuously narrowing the proposal region as optimization progresses. 
We validate across diverse inverse problems from path tracing, vector splines, Voronoi shaders, and robotics, and demonstrate that feature-guiding substantially improves convergence over scalar-loss baselines and reliably escapes local minima where gradient-based methods stall.
\end{abstract}

\begin{teaserfigure}\centering
\includegraphics{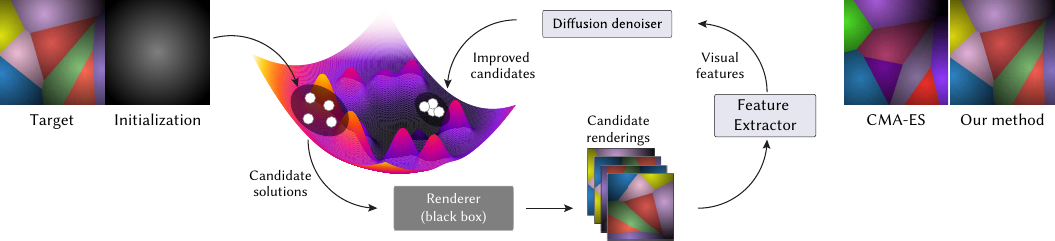}
\caption{Feature-guided diffusion in parameter space. Starting from a black-box optimization problem and an initial candidate population, our method samples candidate solutions, renders their respective outcome images, and --- instead of simply assigning a scalar cost or fitness value to each --- computes their visual features, which are used to train a diffusion denoiser to generate improved candidates in the next iteration.} %
\label{fig:teaser}
\end{teaserfigure}

\maketitle
\fancyfoot[C]{} 
\fancyfoot[L]{} 
\fancyfoot[R]{} 

\section{Introduction}
\label{sec:intro}

A vast amount of scientific research is dedicated to inferring causes from effects, since many practical problems can be formulated as inverse optimization.
In graphics, the dominant approach in recent years has been \emph{analysis-by-synthesis} powered by differentiable renderers~\cite{li2018differentiable, Mitsuba3, nimier2020radiative, laine2020modular, li2020differentiable}, which propagate gradients through the image formation process to update parameters via gradient descent.

While powerful, this approach faces two fundamental challenges.
\textbf{First}, making a renderer differentiable is rarely straightforward: obtaining unbiased derivatives in the presence of discontinuities and visibility changes typically requires careful mathematical derivations and problem-specific algorithms.
The techniques used to differentiate vector graphics rasterization~\cite{li2020differentiable}, for instance, differ markedly from those used in light transport simulation~\cite{li2018differentiable,Vicini2022sdf,Zhang2023Projective}.
\textbf{Second}, even when correct gradients are available, the resulting optimization landscapes are often riddled with local minima, and gradient descent -- a fundamentally local procedure -- routinely fails to find the global optimum~\cite{fischer2023plateau, zhang2025many}.
Even a correctly differentiated renderer is therefore not always sufficient for solving inverse rendering problems reliably.

A growing body of work side-steps the first challenge by relaxing the strict unbiased differentiability requirement and instead estimating or perturbing gradients \emph{stochastically}~\cite{fischer2023plateau, fischer2024zerograds, deliot2024transforming, wang2025stochastic, zhang2025many}.
We share this perspective but take it one step further: we propose a framework that solves the inverse problem in a fully black-box manner, treating the renderer as an arbitrary function whose only requirement is that it produces images.
Importantly, our method never explicitly calculates or estimates gradients of the rendering function%
, and requires no special initialization, while still being able to converge to a global minimum, even in the presence of many local minima. 
Our approach, \textbf{Feature-Informed Diffusion Evolution (FIDE)}, achieves this through two complementary principles: 

\paragraph{Principle 1: Feature guiding.}
The obvious difference between gra\-dient-based methods and black-box optimizers is that the former have access to per-pixel derivatives, which provide the optimizer with spatial awareness of how to improve the current solution. 
Black-box optimizers, on the contrary, have no such luxury;
methods like CMA-ES~\cite{hansen2006cma} or Bayesian optimization (Sec.~\ref{sec:relatedwork}) have to base all of their decisions on a single number per candidate, the value of the loss function.
This is a central limitation we address: disregarding the image information and collapsing the current rendered solution to a single scalar value does not tell the optimizer \emph{what} is wrong with a rendered image or \emph{where} the error lies. 

Our key idea is to overcome this information bottleneck by introducing \emph{feature~guiding}.
Instead of merely reducing each rendered image to a scalar acquisition score, we extract dense visual features using a frozen feature extractor (in our case, DINOv3~\cite{simeoni2025dinov3}) to learn a mapping between visual features and the parameters that produced them.
Then, the features of the target image guide the proposal of new solutions.
As graphics problems produce structured image-domain output, this observation is particularly natural here -- and, to our knowledge, has not yet been exploited by existing black-box optimization methods (Sec.~\ref{sec:relatedwork}).
We emphasize that our use of image features to propose candidate solutions is complementary to the use of neural features as part of the loss function (as in, e.g., LPIPS).

\begin{figure}[b]
    \centering
    \includegraphics[scale=0.5]{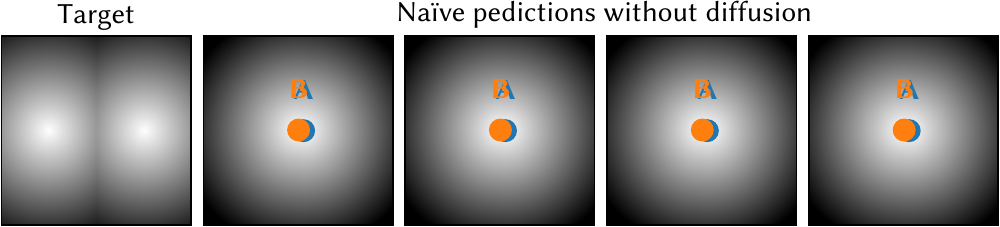}\vspace{0ex}
    \includegraphics[scale=0.5]{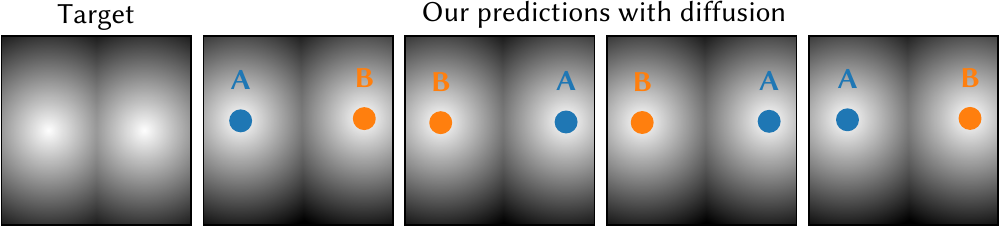}
    \caption{\label{fig:voronoi_2cells}%
        Inferring the centers A and B of two Voronoi cells from an image of their distance transform is ambiguous: the labels are interchangeable.
        A na\"ive predictor (top row) suffers from \emph{mode averaging}, as both cell coordinates are numerically averaged.
        Our predictor is multi-modal using a diffusion model (bottom row); sampling from it yields both global minima.}
\end{figure}

\paragraph{Principle 2: Handling ambiguity through diffusion.}
Additionally, inverse rendering is a frequently ill-posed task, since
the forward mapping is non-injective: many parameter configurations can produce identical or perceptually indistinguishable images (\cref{fig:voronoi_2cells}).
Accordingly, a candidate proposer trained to regress parameters directly from images or visual features
will suffer from two well-known problems: \emph{mode averaging}, i.e., returning the numerical average of the several possible candidate solutions~\cite{bishop1994mixture}, and \emph{mode collapse}, a loss of diversity across stochastic proposals, i.e., concentrating on the same candidate~\cite{salimans2016improved}.

To alleviate these issues, we approximate the inverse function as a multi-modal, conditional distribution using a diffusion model, and, based on the target image features, use it to propose \emph{several} candidates.
The candidates from the diffusion process are subsequently refined through an evolutionary optimization step (\cref{subsec:evolution_step}).
This combination of diffusion and evolution addresses both challenges raised above: maintaining a population of candidates that evolves through stochastic search lets us escape the local minima that trap gradient-based methods, while the diffusion-based proposer lets us represent the inherent ambiguity of the inverse problem and recover valid solutions -- something neither a one-shot inverse predictor nor a generic black-box optimizer achieve on their own. %

\paragraph{Contributions} 
In summary, we make two contributions. First, we introduce a novel black-box inverse-rendering framework that couples a learned, diffusion-based candidate proposer with population-based optimization and requires neither gradients nor initialization. 
Second, we present a thorough empirical study across diverse graphics problems — camera pose estimation, robot arm estimation, path tracing, vector splines, and Voronoi shaders — and show that feature-guiding substantially improves convergence over scalar-loss baselines and that our method remains robust to far-from-optimum initialization and using various loss functions.

\section{Related work}
\label{sec:relatedwork}

\begin{figure}[b]
    \centering
    \includegraphics[width=\linewidth]{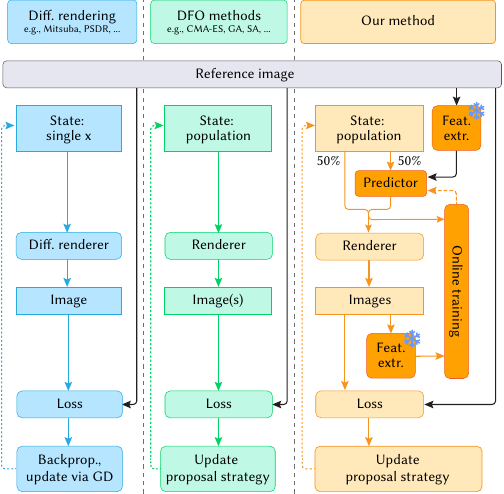}
    \caption{Different optimization methods.
    Black arrows denote the control flow of the target image or its features.
    The feature extractor is frozen.}
    \label{fig:optimization_methods}
\end{figure}

Our work draws inspiration from several lines of research in inverse rendering and numerical optimization, each addressed below; the relationship is illustrated in \cref{fig:optimization_methods}.

\paragraph{Gradient-based inverse rendering.}
Inverse rendering often relies on differentiating the image formation process, either exactly or approximately, to obtain gradients with respect to scene parameters.
Exact differentiation is non-trivial, primarily due to %
discontinuities in the image formation process.
A substantial line of research has addressed this for path tracing through edge sampling~\cite{li2018differentiable}, reparameterization~\cite{loubet2019reparameterizing, bangaru2020warpedsampling, xu2023warped}, and specialized formulations for signed distance fields~\cite{Vicini2022sdf}, while rasterization-based pipelines pursue different strategies based on prefiltering~\cite{li2020differentiable}, anti-aliasing~\cite{laine2020modular}, or specialized shading languages~\cite{bangaru2023slang}.
Other works focused on time and memory efficiency in differentiable light transport, such as radiative backpropagation~\cite{nimier2020radiative, Vicini2021PathReplay}.
The \emph{Mitsuba~3} rendering framework~\cite{Mitsuba3} provides a widely-used implementation of such techniques and is an example of the first column in \cref{fig:optimization_methods}. 

A different class of methods relaxes the strict differentiability criterion and instead aims to \emph{approximate} gradients.
A classic example is the finite difference estimator, where perturbed function evaluations allow defining a derivative, and its multi-dimensional extension SPSA~\cite{spall2002multivariate}.
More recent examples include PRDPT~\cite{fischer2023plateau} with a Gaussian convolution that reduces plateaus in the loss landscape, which was later extended to second order~\cite{wang2025stochastic}; ZeroGrads~\cite{fischer2024zerograds} with a differentiable neural surrogate; 
and a per-pixel stochastic gradient estimation~\cite{deliot2024transforming}.
While these algorithms do not require differentiability, they still perform \emph{gradient-based} iterative optimization.
Our proposed method, in contrast, does not compute or estimate gradients, but instead combines a feature-guided candidate proposal with an evolutionary search in the parameter space.

\paragraph{Derivative-free optimization.}
In contrast to gradient-based methods, derivative-free optimization (DFO), also known as \emph{black-box optimization}, operates purely through function evaluations and does not rely on any gradient information.
As DFO is a wide and mature field, we limit our discussion to algorithms relevant to our approach and refer the interested reader to the comprehensive overview by \citet{conn2009introduction}.
DFO algorithms are often classified as either search-based or evolution-based.
Search-based methods explore the parameter space stochastically without assuming locally coherent behavior.
A well-known example is \emph{simulated annealing}~\cite{kirkpatrick1983optimization}, 
which accepts worsening of the loss, up to an acceptance threshold that decreases
with a cooling schedule, and has been successfully used in inverse
reflector design~\cite{weyrich2009fabricating}.
\emph{Bayesian optimization} fits a probabilistic surrogate -- typically a Gaussian process -- and selects candidates by maximizing an acquisition function; AutoInverse~\cite{ansari2022autoinverse} similarly builds surrogates and focuses sampling on uncertain regions.

Evolution-based methods, in contrast, maintain a population of candidates that coherently evolves over time as it explores the search space.
Examples include \emph{differential evolution}~\cite{storn1997differential} and \emph{genetic algorithms}~\cite{holland1973genetic}, where the population procreates according to a quality measure, and \emph{particle swarm optimization}~\cite{kennedy1995particle}, which defines population motion via a velocity field.
CMA-ES~\cite{hansen2006cma} additionally adapts the search distribution by maintaining a multivariate Gaussian whose covariance is updated from successful samples.
A related line of work augments evolutionary strategies with surrogate models~\cite{loshchilov2013bi} to reduce the number of expensive function evaluations.
When applied to problems in graphics, all of these methods share the limitation we identified in the introduction: their decisions are based on a single scalar fitness value per candidate, discarding the spatial information present in rendered images.
Our method extends the family of evolution-based algorithms with a learned, conditional generative proposer that consumes dense image features rather than the loss function scalars.

\paragraph{Direct inverse prediction.}
Instead of performing iterative optimization, a complementary class of methods learns a direct mapping from images to scene parameters in a single forward pass.
Recent examples include intrinsic decomposition~\cite{Zeng2024RGBXID, Liang2025DiffusionRendererNI} and large reconstruction models~\cite{hong2024lrm,li2025lirm}.
This paradigm has two major limitations.
First, training such models relies on large, representative datasets and an extensive offline training stage that requires millions of images. %
This is necessary since the model must learn the task globally, e.g., the whole domain of 3D objects.
In contrast, our method works adaptively, in an online fashion, fitting an observation with a limited number of training samples/rendering calls and without a dataset (\cref{sec:results}). 
Second, direct prediction methods produce their answer in a single pass through the network, without rendering again to check whether the predicted parameters actually reproduce the target image.
There is no mechanism to refine the prediction against a specific target or recover from a wrong prediction.
In contrast, our method trains its proposer online during the optimization of a \emph{specific} target image; the proposer's outputs serve as \emph{candidates} that are subsequently verified and refined by a population-based evolution.

\paragraph{Diffusion-based optimization}
Adjacent to our method, recent work has explored diffusion models for black-box optimization.
DDOM~\cite{krishnamoorthy2023diffusion} %
train a conditional diffusion model to map desired objective values to candidate solutions but operates in an offline setting, requiring a pre-collected dataset without providing a mechanism for incorporating new evaluations.
Diff-BBO~\cite{wu2024diff} and DiBO~\cite{yun2025posterior} extend this idea to the online setting.
These methods are conceptually close to Bayesian optimization as they use the notion of uncertainty in the acquisition function to select the next candidates.
In contrast to our method, they are still conditioned only on \emph{scalar loss}, not on dense visual features, and they also suffer from the significant performance overhead induced by learning the uncertainty.
On our experiments in Sec.~\ref{sec:results}, not only does DiBO run for several hours compared to minutes of our method, it still cannot converge to the correct solutions because of the missing feature guidance (see \cref{fig:main_comparison}).

\paragraph{Combining features and optimization.}
Recent work has shown the effectiveness of
large pre-trained networks
as visual descriptors that can be used during optimization~\cite{Johnson2016Perceptual, zhang2018unreasonable}.
AlexNet~\cite{krizhevsky2012imagenet} and VGG~\cite{simonyan2014very} features have been used as part of the loss in numerous learning-based methods, including style transfer~\cite{gatys2015neural}, image-to-image translation~\cite{wang2018high}, neural rendering~\cite{aliev2020neural}, and super-resolution~\cite{yuan2022improving}.
At the same time, using frozen pretrained features as input has proven useful in downstream tasks like Content-Based Image Retrieval~\cite{babenko2014neural}, Zero-Shot Object Detection~\cite{radford2021learning}, Appearance Transfer~\cite{fischer2024nerf} and Anomaly Localization~\cite{ardelean2024high}.
Our proposed method employs both ways of leveraging pretrained networks: DINOv3~\cite{simeoni2025dinov3} features 
\emph{condition}
our diffusion-based proposer, allowing it to reason directly about the spatial and semantic content of the current rendered image. 
In addition, we use LPIPS features as our cost function to avoid common issues of pixel-level objectives and achieve a more robust optimization.

\section{Method}
\label{sec:method}
\newcommand{\xstar}{x^*}
\newcommand{\ystar}{y^*}
\newcommand{\finvhat}{\Tilde{f}_\theta^{-1}}
\newcommand{\xstarhat}{\widehat{x*}}

Our method consists of
two key elements:  %
an \emph{inverse prediction step}, followed by an \emph{evolutionary search step}. 
We now
briefly outline the problem
definition
and then describe each step in more detail. 

\subsection{Problem formulation}

We aim to solve an inverse problem for a given function for which the gradients are assumed to be unknown -- either due to discontinuities or zero-gradients in the forward model, or due to a black-box pipeline. 
Concretely, given a function $f$ and the desired output $\ystar$, the task is to find the $d$-dimensional input parameters $\xstar$ for which $f(\xstar) = \ystar$.
Specific to our work, we assume $f$ is an unknown \textbf{rendering} function, rather than an arbitrary black-box function.
That is, $f \colon \IR^d \to \mathcal{I}$, with $\mathcal{I} \subset \IR^{H\times W\times 3}$ being the space of images.

While there exist many, widely-varying optimization methods, we largely identify two categories that differ in how the goal is formulated: direct inversion and progressive optimization.
Direct inversion methods seek to arrive at $\xstar$ by learning a parametric function $\finvhat \colon I \to \IR^d$, where $\theta$ denotes a set of learnable parameters, that approximates the true inverse $f^{-1}$.
This inverse can, for example, take the form of a neural network, trained with a standard loss (e.g., the mean square error) in the parameter space $\IR^d$, on a dataset of random $\left(x, f(x)\right)$ pairs.
After optimizing such a function, it can be used to directly predict the desired parameters $\tilde{x} = \finvhat(\ystar)$.
This technique characterizes offline / amortized optimization methods such as SVBRDF estimation~\cite{deschaintre2019flexible}%

Progressive optimization methods start from one or more initial candidates $x$ and 
refine them to approach $\xstar$ by minimizing a cost $\mathcal{L}(y)$ defined in the output space $\mathcal{I}$, i.e., finding $\tilde{x} = \arg\min_x \mathcal{L}(f(x))$.
As described in \cref{sec:relatedwork}, this is usually achieved by estimating gradients or using a population-based method.

In this paper, we propose a novel approach that can be seen as a combination of direct inversion and progressive optimization.
That is, our approach performs the optimization by alternating two steps: inverse prediction and evolution. 

\subsection{Inverse prediction step}

A limitation of formulating $\finvhat$ as a direct prediction function is that it is, by definition, unable to accommodate non-injective rendering functions $f$.
A well-known example is the ambiguity between lighting and shading: an image of a diffusely shaded red sphere can either be created by a white light and a sphere with red material, or a white sphere lit by red light; the rendering process collapses the underlying different physical parameters to the same image-space representation.
A direct inverse predictor confronted with such an image must commit to a single output, and since several parameter settings are equally consistent with the observation, it is driven toward the mean of the valid explanations -- a setting that typically explains the image worse than any of the individual modes it averages over (for a didactic example, see \Cref{fig:voronoi_2cells}).
Since non-injectivity is the rule rather than the exception in inverse graphics, we instead formulate $\finvhat$ as a generative model, which represents the full posterior over parameters consistent with the observation and lets us sample several valid explanations rather than collapsing them.

\begin{table}[t]
  \centering
  \caption{Notation used throughout the paper.}
  \vspace{-2ex}
  \label{tab:notation}
  \small
  \setlength{\tabcolsep}{4pt}
  \begin{tabular}{@{}ll@{\hspace{1em}}ll@{}}
    \toprule
    Symbol & Meaning & Symbol & Meaning \\
    \midrule
    $d$              & Problem dimensionality                                       & $\mathcal{L}$           & Cost function \\
    $\mathcal{I}$    & Image space, $\subset \IR^{H \times W \times 3}$      & $\theta$                & Network weights \\
    $f$              & Forward function                                             & $\tilde{f}_\theta^{-1}$ & Learned inverse \\
    $f^{-1}$         & True inverse of $f$                                          & $z$                     & Noisy parameter \\
    $x$              & Parameter vector in $\IR^d$                           & $s$                     & Diffusion timestep \\
    $x^*$            & Optimal parameters                                           & $\sigma(s)$             & Noise std.\,dev.\ at $s$ \\
    $\tilde{x}$      & Predicted parameters                                         & $\mu$                   & CMA-ES distribution mean \\
    $y$              & Image, $y \in \mathcal{I}$                                   & $\Sigma$                & CMA-ES covariance matrix \\
    $y^*$            & Reference image                                              & $k$                     & Population size \\
    \bottomrule
  \end{tabular}
  \vspace{-3ex}
\end{table}

More specifically, we implement $\finvhat(z, y, s) \to \IR^d$ as a denoising diffusion model, where $z \in \IR^d$ is a noisy parameter estimate, $y \in \mathcal{I}$ is the conditioning image, and $s$ represents the strength of the noise (timestep in the diffusion process); the noisy $z$ is assumed to come from a Gaussian distribution of standard deviation $s$ around the true $x$ that produces $y$.
Intuitively, the denoising model uses the noisy $z$ to predict the clean parameters $\tilde{x}$, using $y$ as guidance; the objective is for the predicted $\tilde{x}$ to match $x$.
Such a model can be trained in a standard diffusion model fashion; i.e., we sample random parameter vectors $x \in \IR^d$ and random noise factors $s$. 
To obtain the noisy $z$ we use a $d$-dimensional, multivariate Gaussian distribution $\mathcal{N}(x, \sigma(s)^2)$.
Following \citet{li2025back}, we use the \emph{x-pred} and \emph{x-loss}, so that the objective is simply $\| \finvhat(z, f(x), s) - x\|_2^2$.
After the model is trained, we obtain the $\tilde{x}$-prediction by running the backward diffusion process for any random $x$ \textbf{using the ground truth $\boldsymbol{\ystar}$ as condition}.
We sample from the diffusion model using a first-order Euler solver applied to the reverse-time ODE.

Having access to image information from $y$ allows the model to correlate changes in parameters to changes in rendered output and produce valid solutions accordingly. 
Due to memory constraints, we condition the model not directly on image pixels, but on features of image patches instead, which are extracted by the feature extractor $E$. 
Intuitively, this allows the proposal network to ``see'' the current solutions and propose candidates accordingly.
In \cref{subsec:results}, we show that the model is surprisingly robust to the feature representation and emphasize that an unconditional proposal network does not make sense in an inverse rendering setting, since it would just produce random samples from the parameter space. %

We show that our conditioned denoising formulation is able to solve simple inverse problems in \cref{subsec:proof_inverse}.
However, since this approach aims to solve the inverse problem  \emph{globally}, it becomes exponentially more difficult as the problem dimensionality $d$ increases, and the solution quality quickly degrades. 
Considering that we only want to solve the problem for a single $\ystar$, estimating $f^{-1}$ on the entire domain is not needed.
In fact, it would be sufficient to only train the model on parameters $x$ from the vicinity of $\xstar$ -- however, we of course do not know $\xstar$, and thus cannot constrain the sampling to this region of the parameter space.

\begin{figure}
    \centering
    \includegraphics{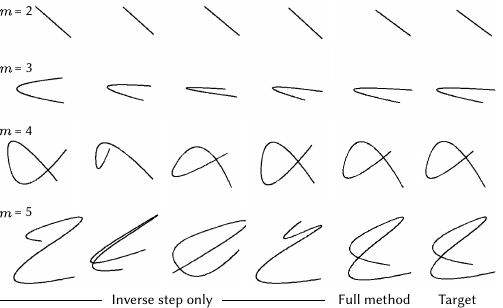}\vspace{-1ex} 
    \caption{\label{fig:just_denoiser}%
    Spline optimization with $m$ control points ($2m$ parameters; rows). The first four columns are outcomes using only our inverse step method (no evolution), followed by our full method and the target image.} %
\end{figure}

\subsection{Evolution step}
\label{subsec:evolution_step}
We therefore introduce our second key idea: \emph{progressive refinement} of our diffusion model's training domain by combining our inverse prediction step with an evolutionary strategy.

Using the candidates proposed by the inverse model, the evolution step explores the parameter space and updates the search distribution based on the achieved loss value of the proposed samples.
In this way, the denoiser provides strong candidates, while the evolutionary loop refines the search space while maintaining exploration and avoiding poor local minima.

We base the evolution step of our proposed two-stage algorithm on CMA-ES~\cite{hansen2006cma}, an established black-box optimizer that has consistently shown good results across different domains. 
However, we note here that our framework is not restricted to this specific evolutionary method, and a different algorithm could be used instead, as demonstrated in the supplement (\cref{asec:our-method-annealing}).

CMA-ES maintains a search distribution $\mathcal{N}(\mu, \Sigma)$ over the parameter space, where the anisotropic covariance $\Sigma$ captures correlations between parameters. %
Each iteration draws $k$ candidates $\{x_i\}_{i=0}^k \sim \mathcal{N}(\mu, \Sigma)$, evaluates their cost $c_i$ according to a pre-defined loss function, and updates $(\mu, \Sigma)$ by computing a weighted average of the candidates, ranked by their cost.
By assigning higher weights to top-scoring candidates during the update step, CMA-ES adapts the search distribution to the local loss landscape.

The synergy between our diffusion-based inverse prediction and CMA is bidirectional.
Firstly, the diffusion model $\finvhat$ estimates a good set of candidates $\smash{\{x_i\}_{i=0}^k}$. 
Having better candidates (closer to $\xstar$) intuitively accelerates the convergence of CMA's covariance matrix toward the desired distribution $\mathcal{N}(\xstar, 0)$.
Secondly, since CMA-ES models the search space as a Gaussian, we can directly use this distribution to train the diffusion denoiser. 
Concretely, we adapt the usual formulation of diffusion models that assume an isotropic noise model, but instead use the anisotropic Gaussian derived by CMA to create training samples. 
Therefore, in our formulation, the timestep scalar $s$ is replaced by the covariance matrix $\Sigma$; in practice, we observe that using the diagonal of the matrix as input to the network is sufficient.
By using the CMA distribution both for training and sampling using the backward diffusion process, we effectively reduce the parameter space which must be modeled by the network.
This reduction is crucial in order for a small network with limited training samples to provide meaningful predictions.

As is customary with optimization methods, finding the right balance between exploration of the search space and exploitation of existing high-ranking candidates is essential for avoiding local minima.
In our approach, this manifests especially in early stages of the optimization, where relying exclusively on the denoising network's predictions to seed the next iteration risks \emph{mode collapse}: the network overfits to its current estimates collapsing prematurely to a single solution. 
To preserve diversity, we apply \emph{partial denoising} --- only half of the CMA candidates are denoised in each iteration, while the remainder is passed through unmodified. 
The full procedure is detailed in Algorithm~\ref{algorithm}.

\newcommand{\algocomment}[1]{\tcp*[r]{\makebox[2.5cm][l]{\parbox[t]{2.5cm}{#1}}}}
\SetKwInput{KwParam}{Parameter}
\begin{algorithm}[t]
    \caption{\label{algorithm} Feature-informed diffusion evolution (FIDE)%
    }
    \SetAlgoLined
    \DontPrintSemicolon
    \KwIn{Renderer $f \colon \IR^d \to \mathcal{I}$, target image $\ystar \in \mathcal{I}$, 
    pretrained feature extractor $E \colon \mathcal{I} \to \IR^{h\times w \times c}$}
    \KwParam{Population size $k$, Iterations $N$, Initial $\sigma_0$}
    \BlankLine
    Initialize CMA-ES: $\mu = \mathbf{0}_d, \Sigma=\sigma_0I_{d\times d}$.\\
    Initialize weights $\theta$ of the denoiser: $\finvhat(z, E(y), \Sigma) \to \IR^d$.\\
    \For{i in $1 \dots N$}{
        Sample $\{x_i\}_{i=0}^k \sim \mathcal{N}(\mu, \Sigma)$ \vspace{0.25em}
        
        $\{x_i\}_{i=0}^{k/2} \gets \finvhat(\{x_i\}_{i=0}^{k/2}, E(\ystar), \Sigma)$ \algocomment{denoise 50\% of samples using target features} %
        $\{y_i\}_{i=0}^{k} \gets f(\{x_i\}_{i=0}^{k})$ \algocomment{render all samples} \vspace{0.25em}
        $\{e_i\}_{i=0}^{k} \gets E(\{y_i\}_{i=0}^{k})$ \algocomment{extract visual features} \vspace{0.25em}
        \tcp{Noise is sampled during training using $\Sigma$}
        Update $\finvhat$ by training $\theta$ using $\{(x_i, e_i)\}_{i=0}^{k}$  \; \vspace{0.25em}
        $\{c_i\}_{i=0}^{k} \gets \mathcal{L}(\{y_i\}_{i=0}^{k})$ \algocomment{compute costs} \vspace{0.25em}
        $\mu, \Sigma \gets \texttt{CMA-ES} (\mu, \Sigma, \{x_i\}_{i=0}^{k}, \{c_i\}_{i=0}^{k})$ \;
    }
    \Return $\mu$
\end{algorithm}

During training (line~8), the diffusion network starts from noisy parameters and is tasked to predict clean parameters conditioned on their renderings, thereby learning a mapping from ``what an image looks like'' to the parameters that produced it. 
During inference (line~5), we condition the model on the target image in order to guide the denoising toward parameters that would render the target. 
In summary, our method combines the advantages of its two components:
The diffusion-based inverse step leverages the rich signal from the target image to generate promising candidates (as opposed to random samples).
The evolutionary step inherits CMA's ability to adapt to the local loss landscape via covariance updates, and, since it is population-based, provides robustness against local minima without explicit gradient information. 
Crucially, the two components reinforce each other — CMA-ES shrinks the region the denoiser must model, while the denoiser accelerates CMA-ES's convergence by seeding it with high-quality candidates.

\begin{figure*}
    \centering
    \includegraphics[scale=0.5]{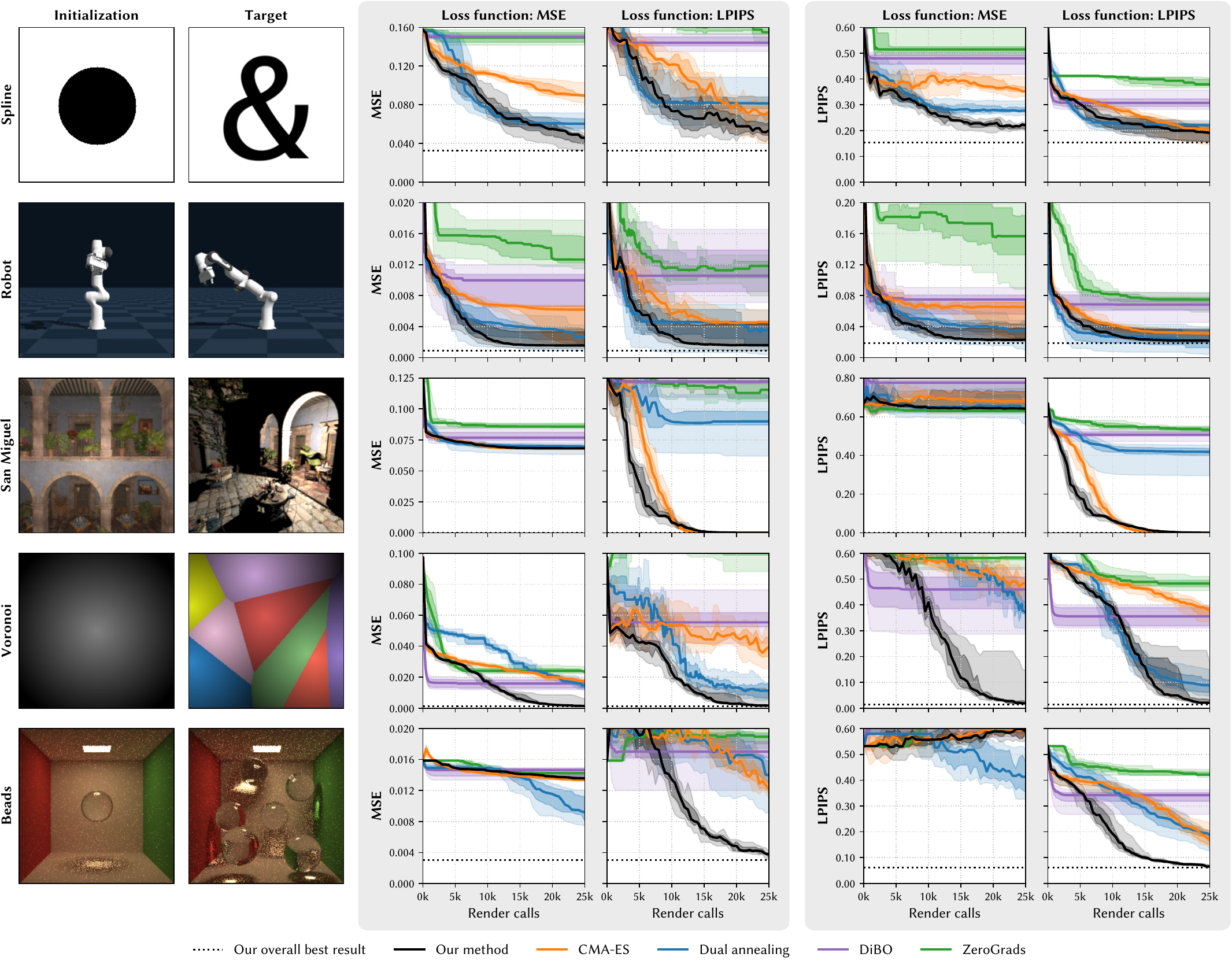}\vspace{-2ex}
    \caption{\label{fig:main_comparison}Comparison of optimization curves on our five test cases.
    We run both our method and the baselines with two different loss functions -- what the optimizer acts on --, namely MSE or LPIPS (columns).
    For each experiment (rows), we additionally cross-report both MSE and LPIPS metrics on the vertical axis, leading to four plots for each test case.
    Each method was executed 10~times with different random seeds and we plot the median (bold line) as well as the approximated 25-75 and 5-95 quantiles (shaded areas).
    Please see the supplement for additional loss function experiments.
    }
\end{figure*}

\subsection{Loss function}
The loss function $\mathcal{L}(y)$ 
can be represented by any image-level cost function that measures how far a certain rendering $y$ is from the target $\ystar$.
Although something like the squared error can be trivially applied, this is generally suboptimal as it is computed at the level of pixels, not taking into account neighboring information.
In \cref{fig:main_comparison}, we show that a perceptual metric, such as LPIPS~\cite{zhang2018unreasonable} generally provides a better optimization objective.
Similarly, \citet{mehta2025locally} have proposed the Locally Orderless Images (LOI) loss to avoid pixel-space problems, on which we evaluate our method in the supplement (\cref{fig:full_report_inclLOI}).
However, rather than advocating for a specific loss function, we argue that robustness across loss functions is itself a desirable property of the optimization method. 
Thanks to our feature-guided diffusion-based candidate proposal, we recover sensible solutions more reliably across different cost functions than previous approaches (see also \cref{fig:main_comparison} and supplement \cref{asec:additional-loss}). 
Additional implementation details are in the supplement (\cref{asec:implementation_details}).

\begin{figure*}
    \centering
    \includegraphics{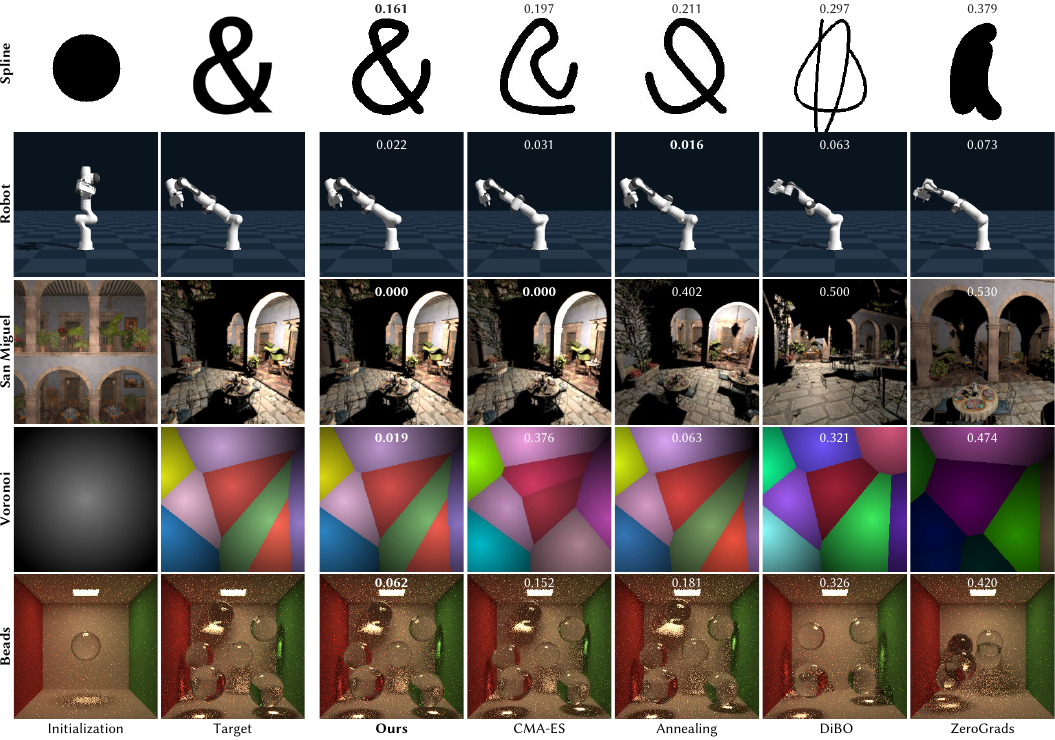}\vspace{-0ex}
    \caption{\label{fig:combined_median}Qualitative analysis. We show the outcome of the median run for our method and the competitors (columns) across all experiments (rows). For visualizations of the best and worst outcome for each method, please see the supplement. Our method achieves a faithful reconstruction in all experiments, even where other methods fail. We display the final LPIPS loss inset in each image, with the best (lowest) outcome per row in bold.}
\end{figure*}

\section{Experiments and results}
\label{sec:results}

\subsection{Validating the inverse step}
\label{subsec:proof_inverse}
Before we evaluate our full method (inverse prediction and evolution), we first
validate our inverse predictor 
on its own
and show the intuition behind using the diffusion model for candidate proposals.
We train the diffusion network globally, that is, on random samples from the entire domain of input parameters $\IR^d$.

The task in this experiment is to optimize the control points of a spline to match a target image.
In \cref{fig:just_denoiser}, we show samples generated by the predictor for the given target images, alongside the results of our full method for reference.
The target images have an increasing number of spline control points (two to five).
Solving the inversion globally is feasible for few parameters, but learning it over the whole domain suffers from the curse of dimensionality.
As \cref{fig:just_denoiser} shows, the inverse predictor achieves plausible approximation of the target for up to four control points, but struggles to generate good solutions when the number of control points increases further (last row).
Crucially,
however, %
even the imperfect samples are closer to the target than purely random draws,
meaning the denoiser still identifies a useful, restricted region of $\IR^d$.

Obtaining such a region significantly diminishes the curse of dimensionality, and a 
same-capacity network with access to this region
will model the local inverse much more accurately. 
This suggests a natural remedy to the global denoiser's degradation: progressively tune the network on samples %
from the neighborhood of its own best predictions. 
In our full method, this neighborhood is precisely what the evolution algorithm tracks and refines -- CMA-ES's search distribution provides both the region to sample training data from and the starting point for the backward diffusion (sampling).

\subsection{Evaluation benchmarks}
\label{subsec:evaluation_setup}

We test our method on five inverse rendering tasks, ranging from robotics to path tracing (\cref{fig:main_comparison}).
Each task is difficult for a different reason, discussed below.
As all methods we compare against are stochastic, they depend on a random seed, so we execute each ten times to account for variations.
In the figures, we report both the median score and quantiles for both our method and the baselines.
Please also see our supplement (Sec.~\ref{asec:implementation_details}) for implementation details.

\paragraph{Spline} We optimize the 2D 
coordinates of seven control points of a quadratic B-spline, together with the spline's thickness (15~parameters in total).
The target image is the \emph{ampersand} symbol created by rendering its font glyph.
This is a hard task since the target image is a filled outline defined by independent Bezier curves, and the optimized spline can only approximate it.
Additionally, the relationship between control points and the resulting curve is non-local: changing a single point deforms the entire curve.
Further, the target has complex topology with a self-intersecting loop, and the order of the control points must be correct.
Finally, the pixel-space loss is indifferent to small position perturbations of points that lie far from the reference, creating plateaus in the loss landscape.

\paragraph{Robot} We optimize the joints of the Franka Emika Panda robot arm~\cite{haddadin2024franka} such that its pose (seven parameters) matches the reference image; a non-trivial 
task: since the mapping from joint angles to image is a composition of highly non-linear forward kinematics under perspective projection.
Additionally, the joints have different effective ranges, leading to a poorly scaled Hessian
(e.g., one joint affects the entire arm's position, while another only changes the topmost segment).
Finally, the
arm covers only a small area of the final image, so the loss metric varies just slightly.

\paragraph{San Miguel} We optimize a camera placement (position and rotation, five parameters) together with a point light position (three parameters).
Both the camera and point light are placed in the San Miguel scene to create a reference image.
The optimizer's task then is to recover all eight parameters, starting at an initial estimate that is far from the optimum.
The loss landscape here is highly non-convex, and the problem is ill-conditioned: small changes in camera and point light positions can lead to big changes in image space (e.g., shadows, visibility behind geometry).

\paragraph{Voronoi} We optimize the parameters of eight Voronoi cells to match a reference image.
Each cell is parametrized by its 2D center position and RGB color, leading to 40~parameters in total.
The difficulty lies in the task's strong permutation symmetry:
swapping any two cells (center and color) leaves the image unchanged, producing $8! = 40{,}320$ global optima.
Further, the Voronoi partition changes discontinuously whenever cell centers swap, making the loss landscape piecewise flat with non-differentiable boundaries.

\paragraph{Beads} This task takes a twist on the classic Cornell box to simulate advanced light transport: we optimize the 3D positions of six glass beads (18~parameters in total) illuminated by an overhead area light.
The caustics, shadows, and reflections within the box create
non-local effects, and the optimizer must operate in the presence of strong Monte Carlo noise.

\paragraph{Baselines}
We compare against four baselines ranging from classical to modern, learned optimization methods.
\textbf{CMA-ES}~\cite{hansen2006cma} serves as our primary evolutionary baseline (see \cref{sec:method} for details).
\textbf{Dual annealing}~\cite{tsallis1996generalized} combines simulated annealing with a local search strategy to improve convergence; we use the standard SciPy implementation \cite{2020SciPy-NMeth}.
From more recent methods, we use \textbf{ZeroGrads}~\cite{fischer2024zerograds}, which trains a differentiable surrogate of the black-box function's loss landscape and uses it to derive optimization gradients; and \textbf{DiBO}~\cite{yun2025posterior}, a diffusion-based method that combines a diffusion prior with a learned reward proxy to sample candidates from an approximate posterior over high-performing solutions.
Note that we do not compare directly to the classical BO, because its performance overhead is non-linear with iterations and several orders of magnitude higher than that of all other methods.

\begin{figure}
        \centering
        \includegraphics[scale=0.5]{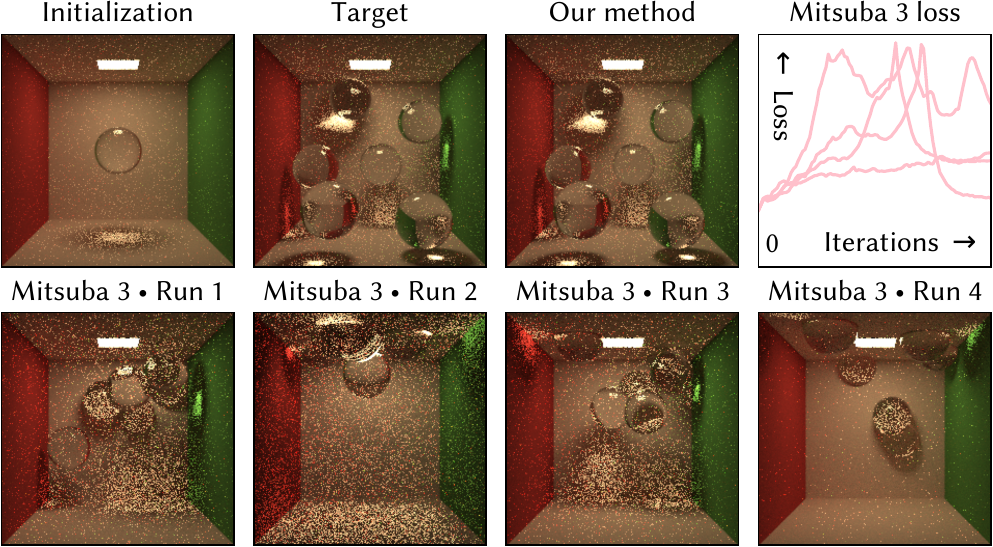}
        \caption{\label{fig:mitsuba}%
            Differentiable rendering is susceptible to diverging if the initial estimate is far from the optimum.
            As an example on our Beads task, we executed the projective sampling integrator (\texttt{prb\_projective}) in Mitsuba\,3 based on \citet{Zhang2023Projective} and it diverged every time.
            In contrast, our method converges even for a suboptimal initializations (\cref{fig:main_comparison}).
            The Monte Carlo noise is left on purpose to demonstrate what all optimizers actually see.}
\end{figure}

\begin{figure}
        \centering
        \includegraphics[scale=0.5]{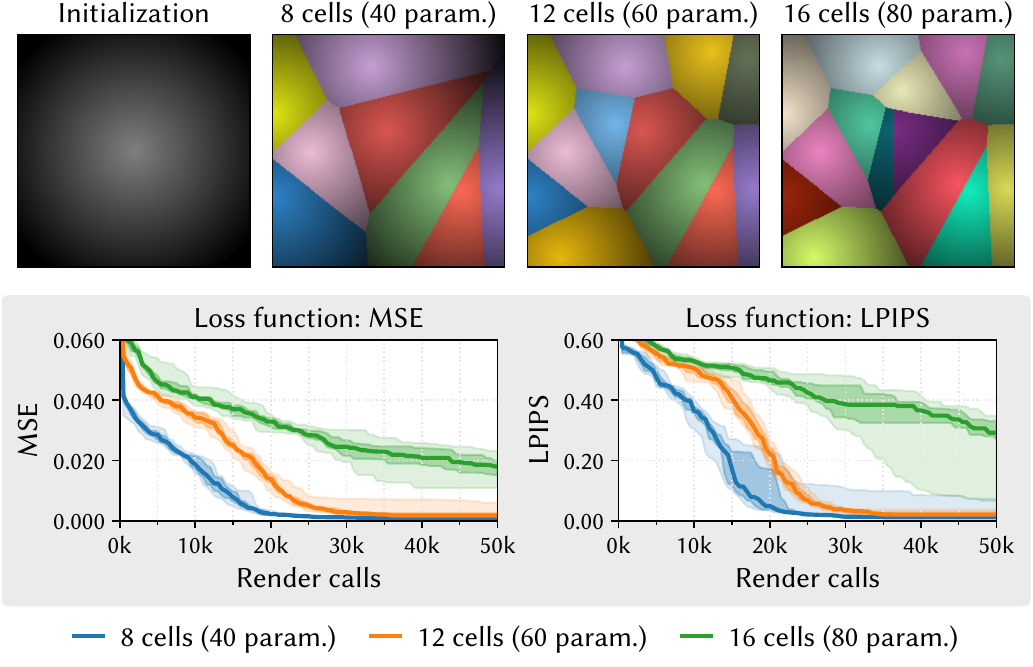}\vspace{-2ex}
        \captionof{figure}{\label{fig:voronoi_dim}%
            Demonstration of our method with a varying number of optimized parameters on the Voronoi task.
            The initialization remains the same, but the target images have an increasing number of cells.
            It is evident that more iterations are required for higher-dimensional problems.}
\end{figure}

\subsection{Results}
\label{subsec:results}

\paragraph{Main quantitative evaluation}
\Cref{fig:main_comparison} presents the main quantitative evaluation, comparing our method to the baselines across our five test cases.
Under the LPIPS loss, 
our approach consistently finds a better solution, and it is less prone to getting stuck in local minima.
Even though on some test cases dual annealing or CMA-ES find solutions
on par in
metrics, 
ours is the only method that recovers them reliably across all tasks.
The observation remains true under the MSE loss, except on the \emph{San Miguel} and \emph{Beads} experiments, where all approaches uniformly fail to obtain a meaningful result for MSE.
We also tested all the optimization algorithms on the locally-orderless images (LOI) loss function based on \citet{mehta2025locally}, which is discussed further in the supplement (\cref{asec:additional-loss}).

\begin{figure*}
    \centering
    \includegraphics[scale=0.5]{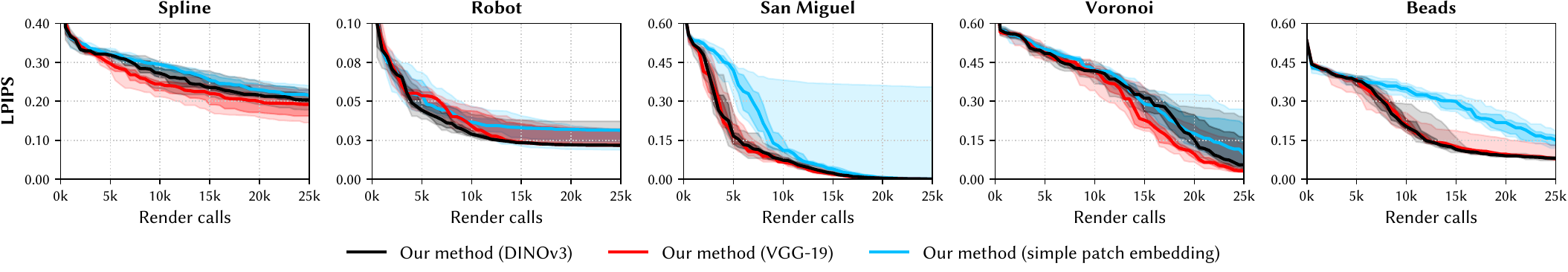}\vspace{-2ex}
    \caption{\label{fig:backbone_comparison}Comparison of optimization results using our method with different visual feature extractors.
    It is evident that our method is compatible with a wide range of features, but pretrained networks such as VGG-19 and DINOv3 improve convergence over a simple patch embedding, especially on images with Monte Carlo rendering noise (San Miguel, Beads).}
\end{figure*}

\paragraph{Qualitative results}
\Cref{fig:combined_median} shows the final outcome of all methods qualitatively.
To avoid judging a method by outliers, we report the result of a middle run -- 4th out of 10 -- in sorted LPIPS order.
By visual assessment, our approach is the only one that reliably solves all five test cases.
Further comparisons and a video sequence of our method's convergence are available in the supplementary materials.

\paragraph{Comparison to differentiable rendering}
In \cref{fig:mitsuba}, we compare our method with Mitsuba\,3.
Using the \texttt{prb\_projective} integrator, Mitsuba computes an unbiased estimate of the gradients, which can be directly used to optimize the parameters; however, due to the gradients being local and noisy, the optimization fails in the presence of plateaus, as the initial estimate is far from the solution. 
We use the default initialization that all experiments see; specialized scheduling and multi-resolution optimization could likely improve Mitsuba's behavior, but the broader point — that gradient-based methods require informed initialization — remains.
In contrast, our approach is able to contend with both noise and unfavorable initialization.

\paragraph{Scaling with dimensionality}
We run the Voronoi task in three variants, at 40, 60, and 80 parameters. 
The per-iteration runtime stays roughly constant across all three settings. 
The number of render calls required to converge, however, grows non-linearly with dimensionality: convergence is fast up to 60 parameters; however reaching a comparable loss at 80 parameters would require more than 50,000 render calls (\cref{fig:voronoi_dim}). This scaling behavior is also task-dependent: \cref{fig:main_comparison} shows that different problems converge at very different rates even at similar dimensionality.

\begin{figure}
    \centering
    \includegraphics[scale=0.5]{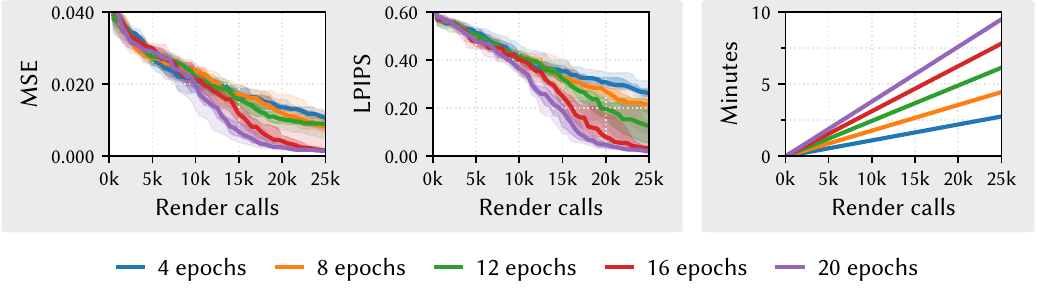}\vspace{-2ex}
    \caption{\label{fig:epochs}%
        Increasing the number of epochs in our predictor training improves convergence, as shown here on the Voronoi test case, but comes at a higher performance overhead and is better suited for expensive black-box functions.}
\end{figure}

\paragraph{Feature extractors}
In \cref{fig:backbone_comparison}, we compare the effectiveness of DINOv3 and VGG-19 as feature extractors.
Both
perform well on average, and we keep DINOv3 (small) as our default 
for
its speed.
\Cref{fig:backbone_comparison} also includes a ``simple patch embedding'', which uses a learnable, one-layer patch embedding (16\texttimes{}16) instead of a pretrained feature extractor.
Note that it would not make sense to run our method completely without any image features, since that would %
yield an unconditional diffusion model sampling
from the parameter space randomly.
While the performance decreases without pretrained features, we observe a graceful deterioration, showing that our method is robust to the type of visual features used.

\subsection{Limitations}
Our approach for optimizing black-box rendering functions combines an inverse-prediction model with a CMA-based evolutionary strategy
and inherits their shared strength as well as limitations.
Due to the quadratic memory complexity of CMA-ES, our method is best suited for small-to-medium sized optimization problems (up to 100 parameters). 
This is reinforced by the
diffusion model's architecture,
which includes quadratic self-attention over the parameters, and poses an interesting question for future research.

\paragraph{Runtime overhead}
Additionally, our image-informed method improves convergence w.r.t.\ the number of rendering calls.
This comes at the cost of computational overhead for 
online predictor training, the latter of which varies with training epochs.
In \cref{fig:epochs}, we show the convergence and total runtime (including render time) for various numbers of training epochs on the Voronoi task.
Further, for the Beads task with 25,000~render calls at 224\texttimes{}224~pixels and LPIPS as the loss, the full CMA-ES execution takes about 4.5~minutes on a desktop computer, while our method takes between 6 to 13~minutes (4 to 20~epochs, respectively), and DiBO about 4~hours.
Naturally, for black-box functions that are more expensive to compute, our method's overhead becomes proportionally smaller.
More importantly, the baselines' lower overhead is irrelevant when they get stuck in poor local optima or fail to converge, as \cref{fig:main_comparison} shows.

\section{Conclusion}
We introduce a novel approach to derivative-free optimization, specifically tailored to black-box rendering functions.
Our method brings together the benefits of inverse prediction and evolutionary strategies by using a diffusion model conditioned on image features to propose candidates, and leveraging the CMA-ES algorithm to iteratively refine the search space.
As shown by our experiments, the mutual synergy between the two components allows our algorithm to navigate local minima more robustly and find better solutions compared to the baselines, using the same number of calls to the rendering function.

\begin{acks}
The authors gratefully acknowledge the scientific support and HPC resources provided by the Erlangen National High Performance Computing Center (NHR@FAU) of Friedrich-Alexander-Uni\-ver\-si\-tät Erlangen-Nürnberg (FAU). The hardware is partially funded by the German Research Foundation (DFG).
\end{acks}

\bibliographystyle{ACM-Reference-Format}
\bibliography{paper}

\clearpage%
\appendix
\setcounter{page}{1}
\setcounter{figure}{0}
\makeatletter
\renewcommand{\thesection}{S\arabic{section}}
\renewcommand{\thefigure}{S\arabic{figure}}
\renewcommand{\thetable}{S\arabic{table}}
\renewcommand{\theequation}{S\arabic{equation}}
\makeatother

\twocolumn[\sffamily%
  \leftline{\Huge Feature-Guided Diffusion for Non-Differentiable Inverse Rendering}
  \bigskip\medskip
  
  \leftline{\LARGE\uppercase{Supplementary Material}}
  
  \bigskip\bigskip%
]

\noindent
This document accompanies the main paper
\emph{Feature-Guided Diffusion for Non-Differentiable Inverse Rendering}
and contains supplemental material referenced therein.

\section{Training and implementation details}
\label{asec:implementation_details}
Our diffusion model is implemented as a transformer-based parameter refinement architecture, conditioned by image features using cross attention.
Firstly, the image features are flattened into a sequence of tokens, to which we add learned positional embeddings, in order to facilitate working with features either from transformer architectures (e.g., DINOv3) or convolutional networks (e.g., VGG-19).
Secondly, the noisy parameters $z$ are projected into a sequence of tokens using a learned linear layer, which implicitly encodes position information.
Then, the parameter tokens are processed by four self-attention blocks followed by cross-attention with the image tokens.
Finally, the parameter tokens are concatenated and projected back to the problem dimension $d$ using a linear layer.

To inject the information about the timesteps we use Feature-wise Linear Modulation (FiLM)~\cite{perez2018film}.
This conditioning is applied in each attention block.
We leverage the anisotropic noise model from CMA by using a varying timestep for each parameter; specifically, we use the square root of the diagonal of the covariance matrix as the timestep at the end of the diffusion process.
To obtain a candidate from the diffusion model, we use a sample from CMA as a starting point and solve the reverse-time ODE with Euler's method with a very small number of steps. 
In our experiments, we empirically observed that 3 steps were sufficient, which correspond to the manually selected schedule of $[0.3, 0.2, 0.1]$.

As described in \Cref{algorithm}, we train the diffusion network online, in tandem with the evolution strategy.
At each iteration, the population of $k$ samples is rendered using the black-box function $f$, creating a dataset of parameter-image pairs.
The diffusion model is then trained on this dataset, starting from the weights from the previous iteration of the algorithm.
For consistency, we use the same hyperparameters across all five inverse rendering tasks (see \cref{subsec:evaluation_setup}), namely: population size $k=500$, number of evolution cycles $N=50$, initial standard deviation $\sigma=1/6$, and training for 20 epochs using the Adam optimizer with a learning rate of $0.001$.%

\paragraph{Tasks parametrization}
All our tasks have their parameters normalized to values in the $[0,1]$ interval and the initial parameters are always at $0.5$, the middle of the interval.
That also motivates our method's initial standard deviation $\sigma=1/6$, as the entire parameter space $[0,1]$ is within $3\sigma$ of the initialization.
In Spline, the control points are allowed to be anywhere in the image, so 0 and 1 correspond to the image bounds; the maximum thickness is equal to the image size.
In Robot, the $[0,1]$ interval is mapped to the joint limits of the Franka Emika Panda robot.
In San Miguel, the camera and point light positions are parametrized by their XYZ coordinates roughly within the inner scene bounds ($[7,25]$, $[1,10]$, $[-2,12]$), and the camera orientation is parametrized by spherical angles.
In Voronoi, the cell centers are mapped to the image bounds and the colors are mapped to the RGB range.
In Beads, the beads positions are mapped to the bounds of the scene, within the walls.

\paragraph{Tasks implementation}
All our tasks are rendered at 224\texttimes{}224 pixels for performance reasons.
The Spline task is implemented as a simple Python CPU renderer, which interpolates the spline curves using SciPy \cite{2020SciPy-NMeth} and draws the subdivided lines using OpenCV.
For Robot, we use the Genesis library and load the scene using \texttt{genesis.morhps.MJCF} from the library's \texttt{panda.xml} file.
For San Miguel and Beads, we use Mitsuba\,3 in the CUDA mode, with the direct (single bounce) integrator at 8~samples per pixel (spp) for San Miguel, and a path tracer with 64~spp for Beads.
For Voronoi, we use an OpenGL fragment shader.

\paragraph{Baseline hyperparameters}
For our baseline algorithms, we consistently use the same hyperparameters for all tasks.
In CMA-ES, to ensure a fair comparison to our method, we settled on the same population size $k=500$, number of evolution cycles $N=50$, and initial $\sigma=1/6$ as in our method.
Note that $k\cdot{}N=25000$, which is the total number of render calls reported in most our figures.
For dual annealing, we use the default settings of \texttt{scipy.optimize.dual\_annealing} with the maximum number of function evaluation calls set to 25,000.
For ZeroGrads, we consulted the hyperparameters with the authors: we use the defaults in their source codes in the low-dimensional problem regime, set $\sigma=1/6$ following our task parametrization, 1,250~epochs, batch size of~10, and enabled antithetic sampling, which doubles the render calls to a total of 25,000.
For DiBO, we use the hyperparameters from similar-dimensional problems in their source code: batch size~50, 100~initial samples, 50~epochs (proxy, prior, posterior), 30~local search epochs, 30~diffusion steps, and a buffer size of~300; the total evaluations are again capped at 25,000.

\paragraph{Source code}
While we deem the implementation details above to be sufficient to fully replicate our work, we intend to release our source code upon acceptance.

\section{Additional loss metrics}
\label{asec:additional-loss}

In the main paper, \cref{fig:main_comparison} is only a subset of the experiments we ran.
To show the robustness of our method, we further optimized all test cases with an additional loss function based on the locally-orderless images (LOI)~\cite{mehta2025locally}.
Because the LOI optimizer requires a differentiable renderer, which we do not have in the context of our paper, we instead use their histogram-based LOI values directly as a loss function.

\begin{figure*}[p]
    \centering
    \includegraphics[width=\textwidth]{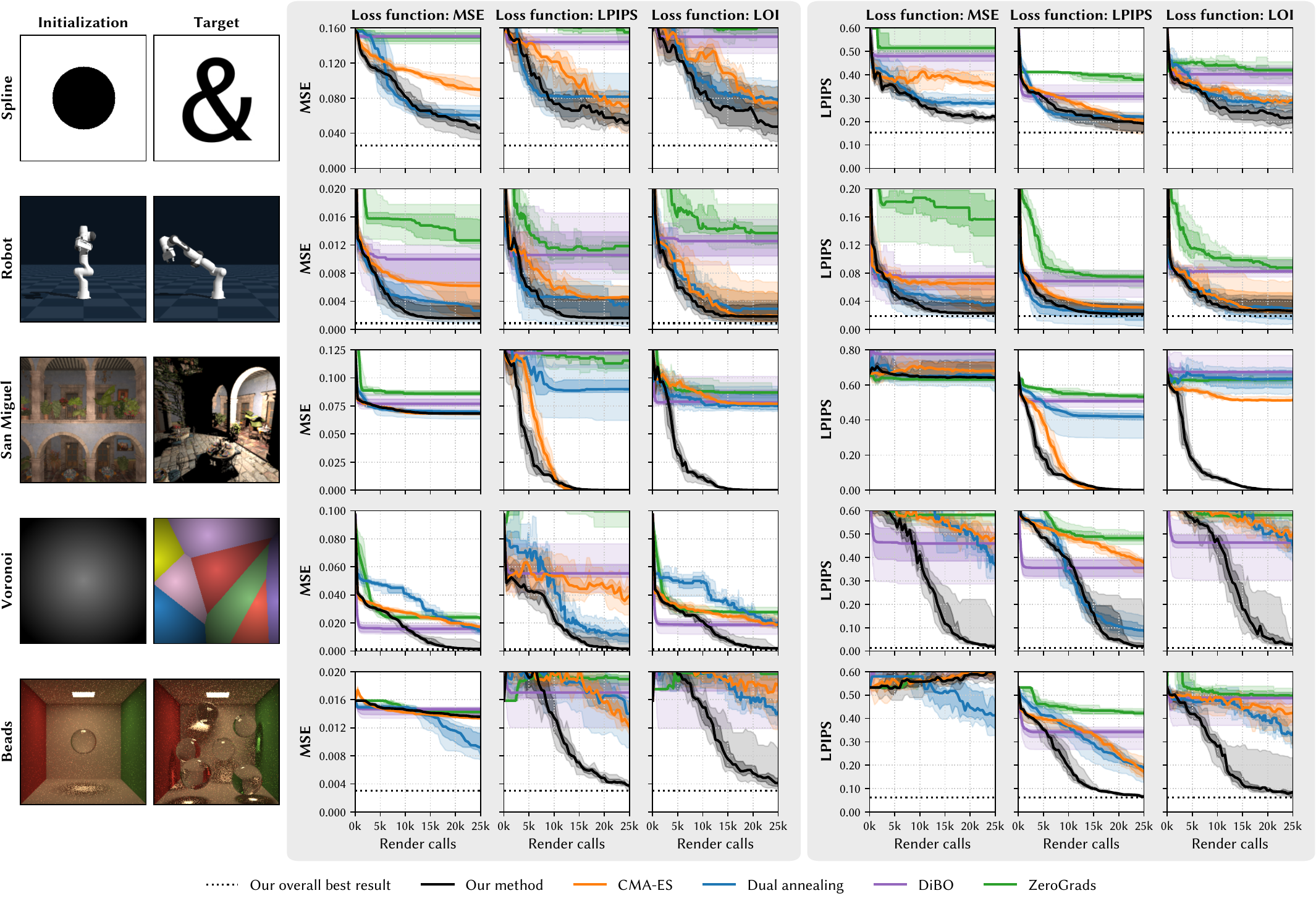}\vspace{-2ex}
    \caption{\label{fig:full_report_inclLOI}Extended version of \cref{fig:main_comparison} in the main text. In addition to it, we plot here the results obtained using LOI as an optimization objective.
    For each experiment (rows), we cross-report both MSE and LPIPS metrics on the vertical axis.
    Each method was executed 10~times with different random seeds and we plot the median (bold line) as well as the approximated 25-75 and 5-95 quantiles (shaded areas).
    }
\end{figure*}

In \cref{fig:full_report_inclLOI}, we show the full report showing the MSE and LPIPS scores while the optimizations themselves used MSE, LPIPS, and LOI as optimization objectives (loss functions).
We observe that our method is more robust to the loss function selection than other methods.
For example, it converged on the San Miguel, Voronoi, and Beads tasks with LOI loss, even though the other methods could not.
However, on our specific tasks, we did not observe that the LOI loss would necessarily yield better results than LPIPS.

We therefore left this part for the supplement, although it serves as an interesting confirmation that our approach is relying both on the extracted visual features \emph{and also} the loss values of the particular loss function to help drive the population to the global optima.
This is further evident from the fact that the San Miguel and Beads tasks, for instance, did not converge under MSE loss for any method, but converged both under LPIPS and LOI with our method.

\section{Extended qualitative comparison}
\label{asec:full-qualitative-comparison}

\begin{figure*}[p]
    \centering
\includegraphics{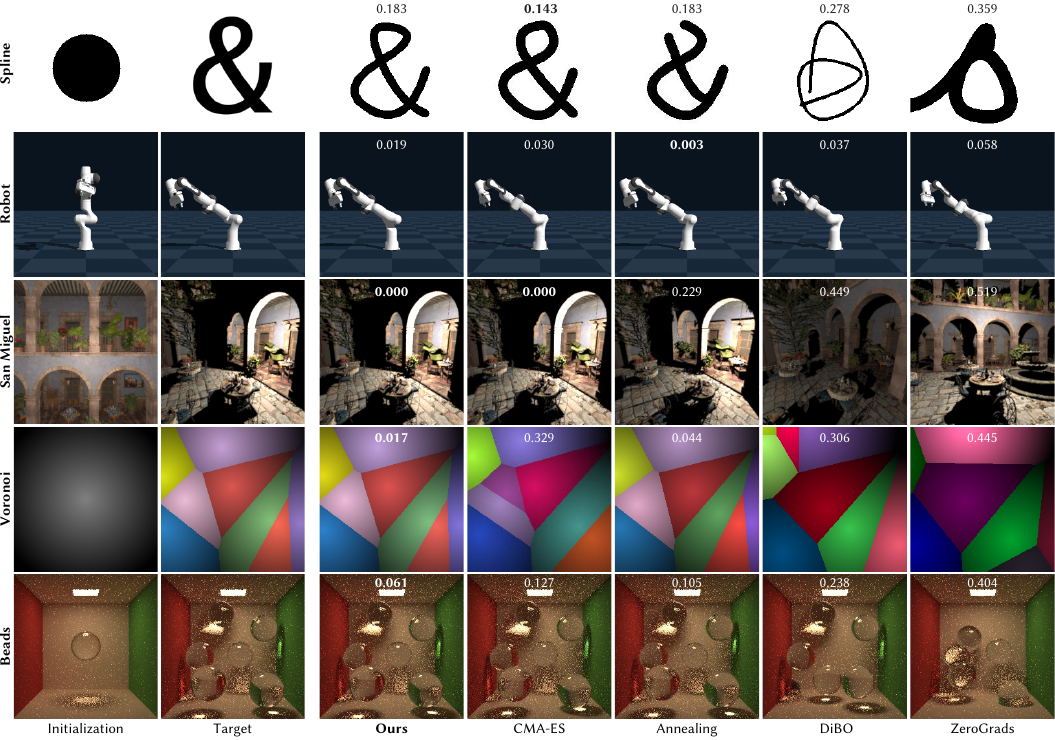}\vspace{-2ex}
    \caption{\label{fig:combined_best}\textbf{Best outcome} of each method's 10 optimization runs across all experiments (rows). We display the final LPIPS loss inset in each image, with the best (lowest LPIPS) outcome per row in bold. 
    }
\end{figure*}

\begin{figure*}[p]
    \centering
\includegraphics{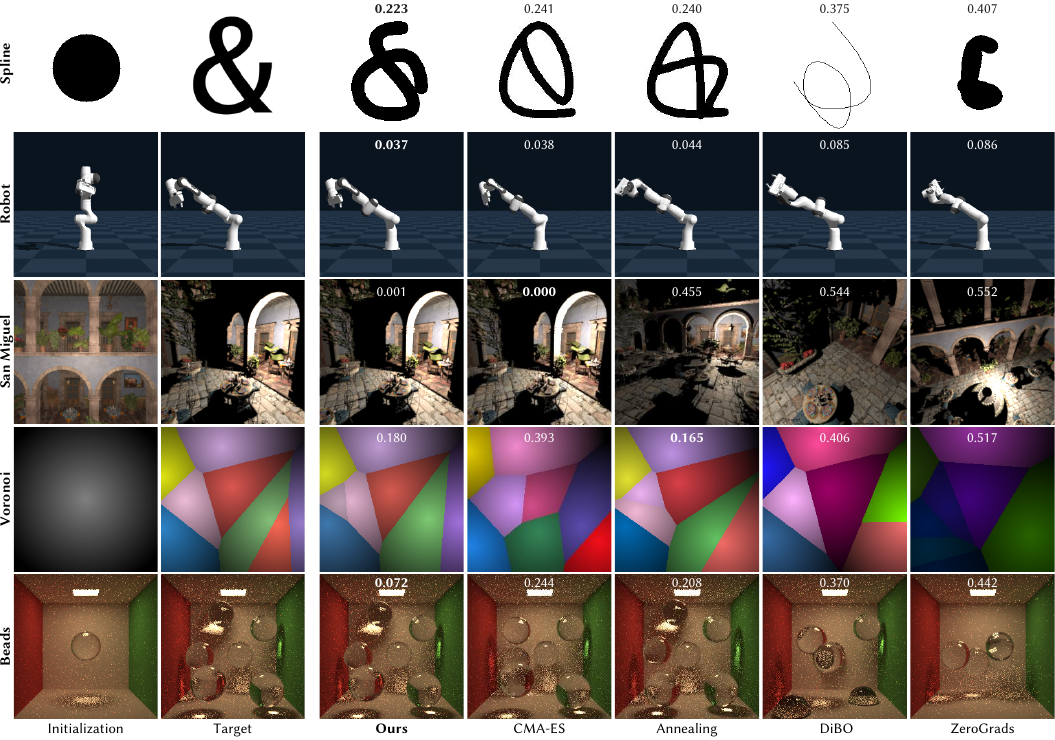}\vspace{-2ex}
    \caption{\label{fig:combined_worst}\textbf{Worst outcome} of each method's 10 optimization runs across all experiments (rows). We display the final LPIPS loss inset in each image, with the best (lowest LPIPS) outcome per row in bold. 
    }
\end{figure*}

In the main paper, \cref{fig:combined_median} shows the images of the median results on our five tasks using all the methods.
As we executed each method 10~times, for completeness, we also include the \emph{best results} (\cref{fig:combined_best}) and \emph{worst results} (\cref{fig:combined_worst}).
The figures confirm that our method is the most robust and consistent.
While the other methods can sometimes succeed with a ``lucky'' random seed driving their stochastic process, our method remains consistent across most of the runs.

\section{Extension to dual annealing}
\label{asec:our-method-annealing}

\begin{figure*}[p]
    \centering
    \includegraphics[width=\textwidth]{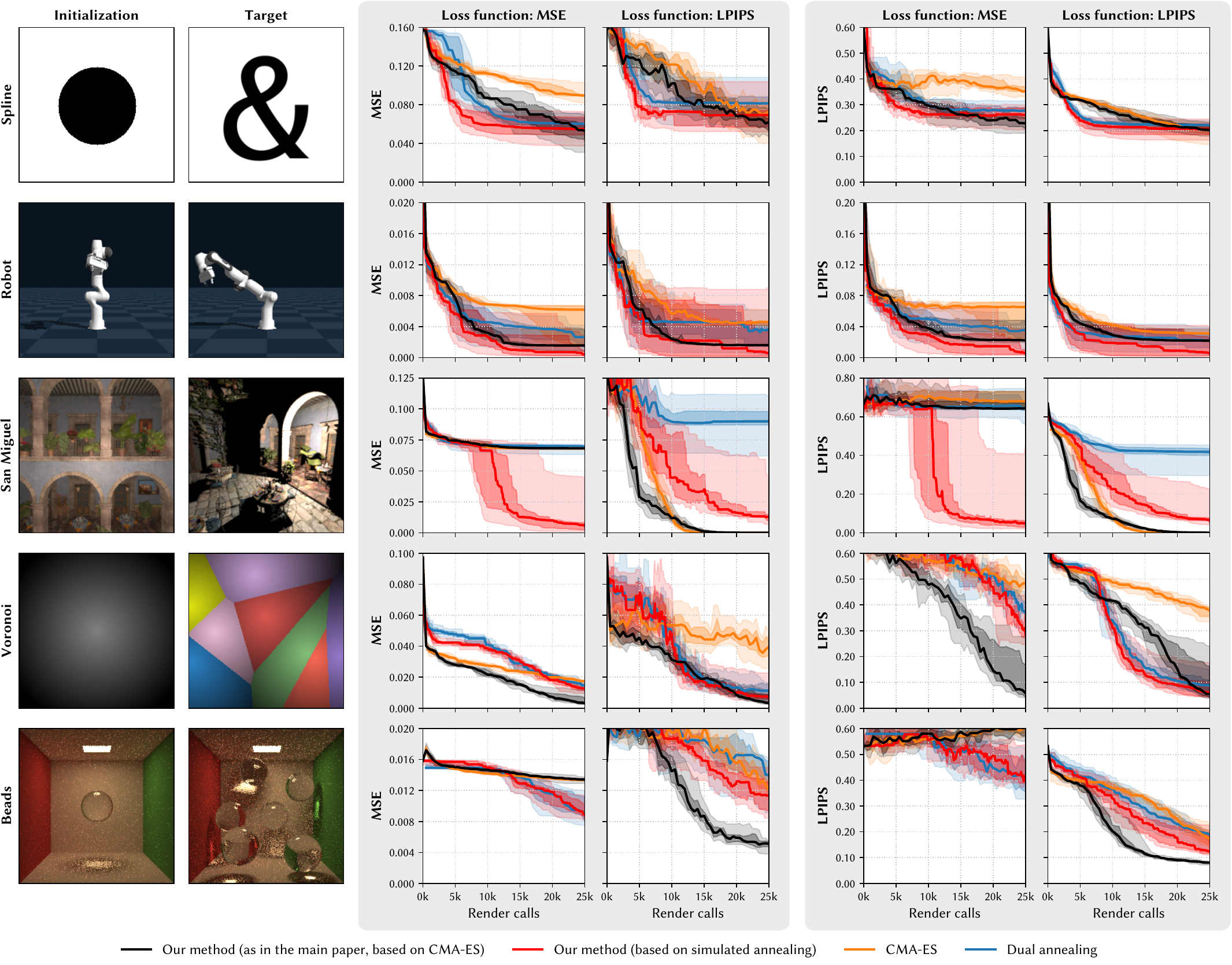}\vspace{-2ex}
    \caption{\label{fig:our-method-annealing}Application of our method to different algorithms, using the same figure format as \cref{fig:main_comparison} in the main text. Here we show that our method can also be executed with dual annealing instead of CMA-ES as the evolutionary step, and that it improves over the standard dual annealing baseline. We also add CMA-ES and our method combined with CMA-ES (as reported in the main paper) for reference. Note that this experiment uses 12 epochs for both ours + annealing and ours + CMA-ES.
    }
\end{figure*}

In the main paper, we present our method using CMA-ES as the underlying mechanism for refining the search distribution.
This coupling is especially fitting due to the explicit representation of this distribution as a covariance matrix in CMA; nevertheless, our approach can be conceptually coupled with other global search heuristics.
To demonstrate this versatility, we also combine our inverse step prediction with dual annealing instead of CMA-ES, and we present these results in \cref{fig:our-method-annealing}.

There are only a few changes relative to \Cref{algorithm}: 
The simulated annealing implementation uses a distorted Cauchy-Lorentz visiting distribution. To account for this, we calculate the covariance matrix empirically using previous samples.
That is, we accumulate both the samples produced by the visiting distributions and the ones denoised using our model up to the population size $k$ (which we set in the same way as for CMA-ES). 
After reaching $k$ samples we train the model for $12$ epochs and continue the simulated annealing procedure. 
Similarly to line 5 in \cref{algorithm}, we always use half of the samples produced by the visiting distribution directly and denoise the other half using our diffusion model.
The accept-reject decision is then applied uniformly to the usual samples and the denoised samples.
The local search part of dual annealing is left unchanged, using L-BFGS-B as per the implementation in Scipy.

The experiment in \cref{fig:our-method-annealing} shows that combining our method with dual annealing consistently outperforms dual annealing by itself.
This suggest the potential of our framework to adapt to different global search heuristics.

\end{document}


\maketitle

\noindent
This document accompanies the main paper
\emph{Feature-Guided Diffusion for Non-Differentiable Inverse Rendering}
and contains supplemental material referenced therein.
References of the form \mbox{Sec.~X}, \mbox{Fig.~X}, etc.\ without an ``S''
prefix point to the main paper.

\section{Full method comparison}
\begin{figure*}[p]
    \centering
    \includegraphics[width=\textwidth]{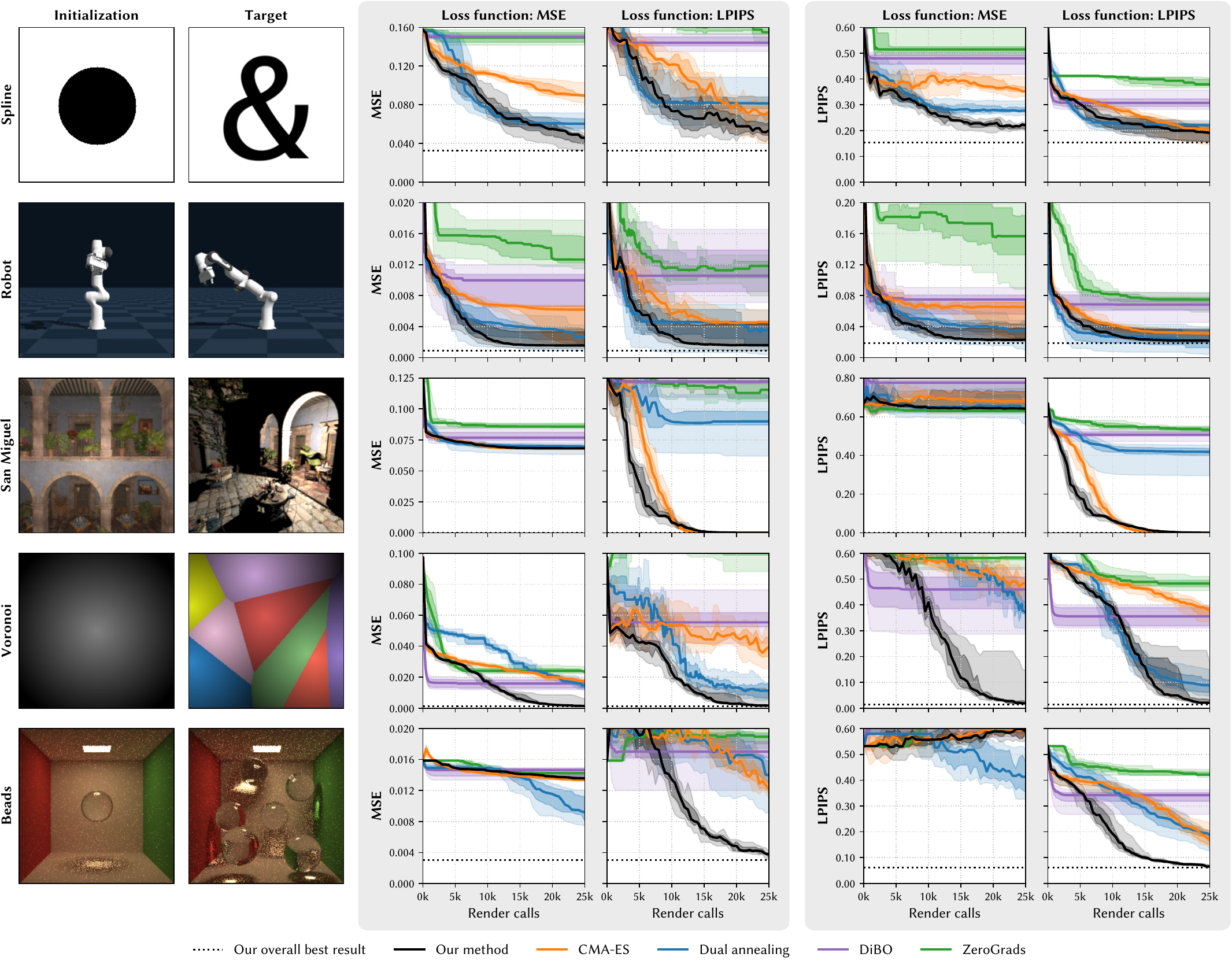}\vspace{-2ex}
    \caption{\label{fig:full_report}Comparison of optimization curves on our five test cases.
    }
\end{figure*}

In the main paper, \cref{M-fig:main_comparison} is only a subset of the experiments we ran.
To show the robustness of our method, we further optimized all test cases with an additional loss function based on the locally-orderless images (LOI) \cite{mehta2025locally}.
Because the LOI optimizer requires a differentiable renderer, which we do not have in the context of our paper, we instead use their histogram-based LOI values directly as a loss function.
In \cref{fig:full_report}, we show the full report showing the MSE and LPIPS scores while the optimizations themselves used MSE, LPIPS, and LOI as optimization objectives (loss functions).

\bibliographystyle{ACM-Reference-Format}
\bibliography{paper}